\documentclass[a4paper]{jpconf}
\usepackage{graphicx}
\usepackage{subfigure}

\begin{document}
\title{Publication patterns in HEP computing}

\author{ M. G. Pia$^{1}$, T. Basaglia$^{2}$, Z. W. Bell$^{3}$, P. V. Dressendorfer$^{4}$}

\address {$^1$INFN Sezione di Genova, 16146 Genova, Italy}
\address{$^2$CERN, 1211 Geneva, Switzerland}
\address{$^3$ORNL, Oak Ridge, TN 37830, USA}
\address{$^3$IEEE, Piscataway, NJ 08854, USA}

\ead{Maria.Grazia.Pia@cern.ch}

\begin{abstract}
An overview of the evolution of computing-oriented publications in high energy
physics following the start of operation of LHC. Quantitative analyses are
illustrated, which document the production of scholarly papers on
computing-related topics by high energy physics experiments and core tools
projects, and the citations they receive. Several scientometric indicators are
analyzed to characterize the role of computing in high energy physics
literature. Distinctive features of software-oriented and hardware-oriented
scholarly publications 
%in the domain of experimental high energy physics 
are highlighted. Current patterns and trends are compared to the situation in
previous generations' experiments.

\end{abstract} 

\section{Introduction}

Publications in scholarly journals establish the body of knowledge deriving from
scientific research; they also play a fundamental role in the career path of
scientists and in the evaluation criteria of funding agencies.

A previous scientometric study \cite{swpub} highlighted that
software-oriented publications are underrepresented with respect to
hardware-oriented ones in the field of high energy physics (HEP).
The results of that analysis showed that the relative
difference between the production of scholarly literature in these areas
had increased in the context of the experiments at LHC (Large
Hadron Collider) with respect to the previous generation's experiments at LEP
(Large Electron-Positron collider).
The analysis in  \cite{swpub} was performed prior to the start of
operation of LHC.

The scientometric analysis summarized in this paper, which reflects a
presentation on this topic at the CHEP (Computing in High Energy Physics) 2012
conference, reviews the publication patterns in HEP computing in greater
detail, with special emphasis on their evolution since the beginning of LHC
operation.

\section{Scope of the study}

The study summarized in this paper provides a quantitative overview of
publication patterns in high energy physics over the past thirty years, with
emphasis on software-oriented publications.

The scientometric analysis is focused on a set of topics that are representative
software R\&D (research and development) in the context of HEP.
The selection is far from exhaustive of the wide variety of research activities
in experimental high energy physics, rather it intends to highlight some
distinctive features of the literary production in the field.

The analysis concerns a representative sample of general software tools,
which respond to common needs of the HEP experimental community, 
and a sample of HEP experiments of the current and past generation.
Two widely used general software tools, Geant4 \cite{g4nim,g4tns} and ROOT
\cite{rootnim,rootcpc} are the object of a detailed scientometric analysis.
More limited investigations concern the publications associated with other
software tools contributing to the general computing infrastructure of LHC
experiments, such as the LHC Computing Grid.
The four major experiments at LEP, the ALICE, ATLAS, CMS, LHCb and TOTEM
experiment at LHC, and the BaBar experiment at the SLAC B-factory are 
included in the scientometric analyis.

The sample subject to evaluation consists of regular publications in established
peer reviewed journals.
Contributions to conference proceedings, books, institutional reports, items in
preprint archives, white papers posted on web sites and software manuals are not
considered.
Some journals (e.g. Nuclear Instruments and Methods, NIM) also publish conference
proceedings, usually in dedicated issues: these articles have been identified and 
excluded from the analysis.

The examined scientometric indicators include the number of publications
produced by the various subjects under study, their time distribution, the
journals where they are published and their citation patterns.

\section{Data sources and analysis methods}

The main source for the scientometric analysis reported in this paper is
Thomson-Reuters' Web of Science \cite{wos}, which is considered the
most authoritative reference for bibliometric information in the academic
environment.
The authors' institutional subscription gives access to a subset of it, the
``Science Citation Index Expanded'' database; it does not include the
``Conference Proceedings Citation Index''.
The database covers the period from 1970 to date.

The access to a subset of the Web of Science generates an apparent mismatch 
between the total number of citations associated with a paper, which includes
entries from the ``Conference Proceedings Citation Index'', and the actual number of
citations available for analysis, which is limited to publications in journals 
belonging to the ``Science Citation Index Expanded''.
A further complication for scientometric analysis is due to the incorrect
classification of some publications listed in the ``Science Citation Index
Expanded'' as ``Conference proceedings'': this label is arbitrarily attributed
by Thomson-Reuters to some regular articles in journals that never publish
conference proceedings (e.g. IEEE Transactions on Nuclear Science, TNS).
Conversely, some entries in the ``Science Citation Index Expanded'' that are not
labeled as ``Conference proceedings'',
appear in journals (e.g. Nuclear Instruments and Methods)
as contributions to conference proceedings.
These errors in the Web of Science have been manually corrected in the analysis
whenever possible: for instance, all papers published in IEEE Transactions on
Nuclear Science are considered in the analysis as regular journal publications,
irrespective of Thomson-Reuters' classification of some of them as conference
proceedings, and papers published in Nuclear Instruments and Methods A issues
dedicated to conference proceedings have been removed from the analysis, even if
they are not identified as ``Conference proceedings'' in the Web of Science.

Other sources have been used to cross-check and complement the
information derived form the Web of Science: the web sites of the 
publishers of technological journals relevant to HEP
and CERN Document System (CDS).
The comparison of the data retrieved from these sources has highlighted some
omissions and inconsistencies in the data sample retrieved from Thomson-Reuters'
Web of Science: for instance, some papers published by LHC experiments, which
are listed in the CDS database, do not appear in the Web of Science, and the
number of citations of a paper reported in the publisher's web site is in some
cases inconsistent with that reported by the Web of Science.

The publications by HEP experiments are distinguished into physics papers (i.e.
publications of experimental results representing the object of the experiment)
and technological papers (i.e. publications about the instruments and methods
that contribute to produce the experimental results).
Technological publications are further identified as hardware-oriented, software
oriented or dealing with data acquisition (DAQ) and trigger.
This classification implies some degree of subjectivity, which has been
mitigated by performing cross-checks over the selections performed by individual
analysts.
It is worthwhile to note that the classification of publications is
part of the regular professional practice of the authors of this paper, either
as members of the Editorial Board of a core journal in nuclear technology or as
responsible of the library of a major HEP laboratory.

The attribution of a paper to a given category is based on a variety of
criteria. 
In some cases the title of a paper or the journal where it is published
unambiguously identify its topic: for instance, papers published in
\textit{Physical Review D} are all classified as physics papers.
For most papers, the record in the Web of Science, which also includes the
abstract, provides sufficient information to identify the scope of the paper and
to classify it in one of the above mentioned categories.
In cases where the attribution is not evident based on the information in the
Web of Science, the full text of the paper was evaluated.

Complementary analyses based on the Web of Science and on publishers' web sites
data, performed independently by different analysts, confirmed the robustness of
the classification.
Based on detailed cross checks over selected samples, the uncertainty in the
results reported in this paper, which derives from intrinsic inconsistencies in
the Web of Science and from subjective classification of papers, can be
estimated of the order of a few percent.
This level of uncertainty does not affect critically the conclusions of this
study.

The analysis reported in the following sections covers three decades of
scientific literature (1982-2011); it is limited to papers published until the
end of 2011 to ensure the reproducibility of results based on the Web of Science.
Unless differently stated, the number of citations reflects the status in the
Web of Science as on 14 May 2012, i.e. one week prior to the beginning of the
CHEP 2012 conference.

\section{General software tools}
\label{sec_tools}
Two software tools used by LHC experiments for simulation and data analysis have
been evaluated: Geant4 and ROOT.

Geant4 is documented in two reference publications \cite{g4nim,g4tns}, which are
brought to the attention of the experimental community in the Geant4 web page.
These papers have collected respectively 2934 and 574 citations (including
citations from conference proceedings indexed by the Web of Science); reference
\cite{g4nim} has crossed the threshold of 3000 citations shortly after the CHEP
conference (3037 citations by 18 June 2012).
Reference \cite{g4nim} is the most cited publication in the "Nuclear Science and
Technology" category over the period considered in this scientometric study.
Excluding the Review of Particle Properties, it is the most cited paper produced
by CERN and by INFN.

\begin{figure}
\centerline{\includegraphics[angle=0,width=12cm]{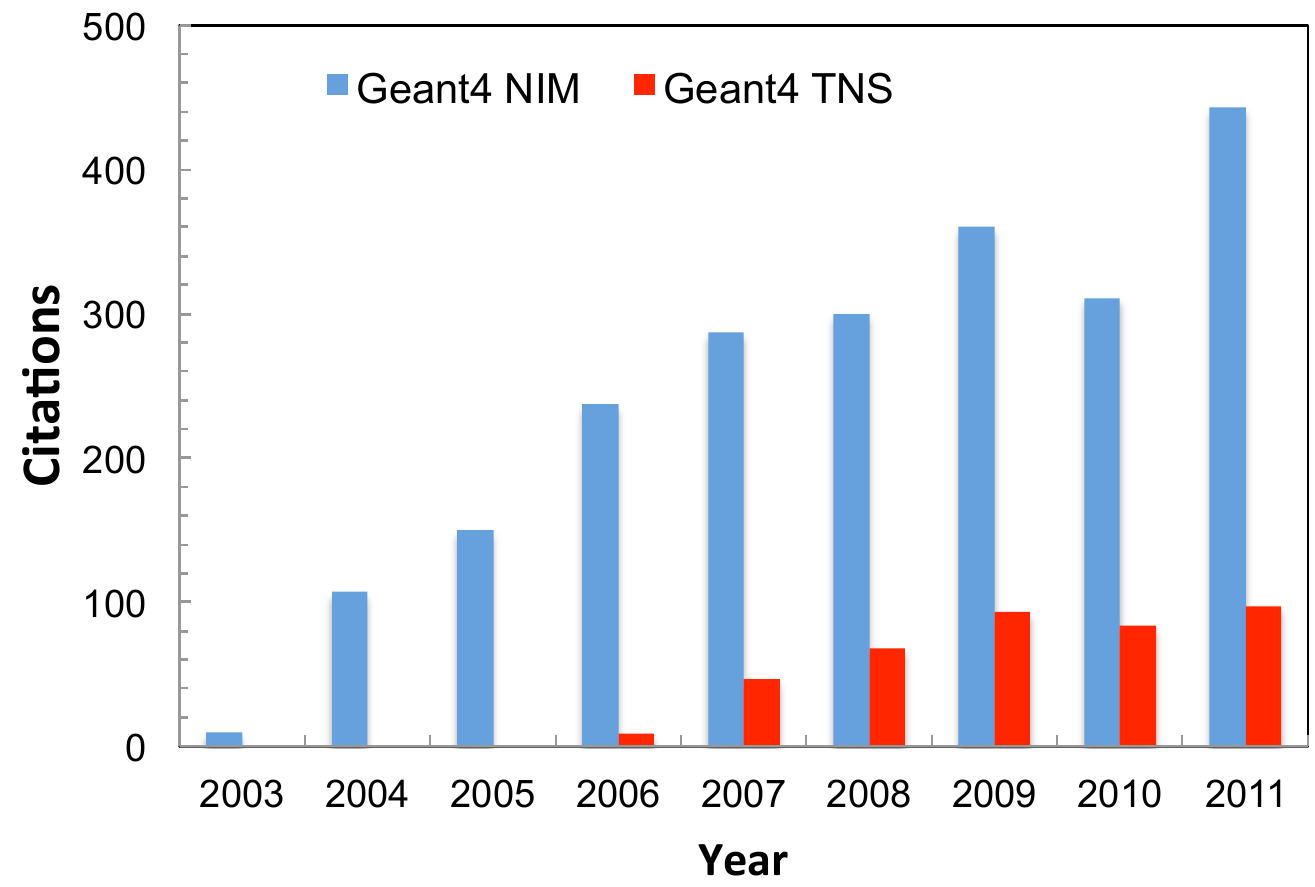}}
\caption{Number of citations collected by Geant4 reference papers
\cite{g4nim,g4tns} as a function of time.}
\label{fig_g4cite_years}
\end{figure}

The time distribution shown in figure \ref{fig_g4cite_years} shows that citation
to the more recent reference \cite{g4tns}, published in 2006, is omitted by most
publications that cite the earlier one \cite{g4nim}, published in 2003.

\begin{figure}
\centerline{\includegraphics[angle=0,width=12cm]{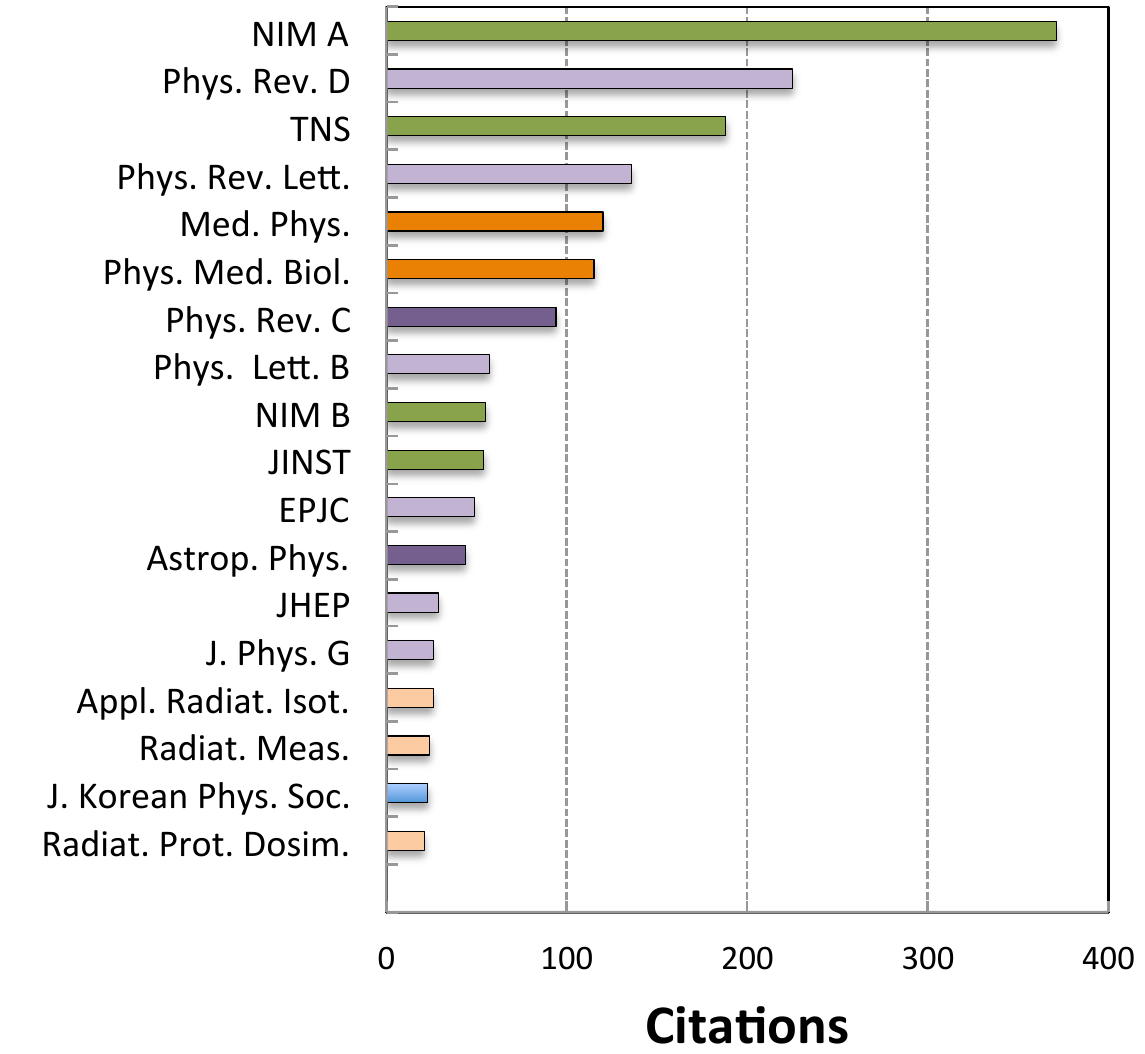}}
\caption{Journals citing Geant4 reference \cite{g4nim}; the colour codes in the
plots are associated with the scope of the journals (physics: violet, nuclear
technology: green, medical physics: orange, radiation protection: light brown,
regional: blue. The journals listed in the histogram contribute approximately 75\% of the
total number of citations to \cite{g4nim}.}
\label{fig_g4cite_journals}
\end{figure}

Although the development of Geant4 was originally motivated by the requirements
of LHC experiments, the source of the citations to its reference paper
\cite{g4nim} shows the widely multidisciplinary character of its use in the
scientific community. Figure \ref{fig_g4cite_journals} lists the journals
contributing the largest number of citations to \cite{g4nim}: it includes
physics journals with various scope (high energy physics, nuclear physics,
astroparticle physics), nuclear technology journals, medical physics and
radiation protection journals, and a regional journal (published by a national
physics society).

\begin{figure}
\centerline{\includegraphics[angle=0,width=12cm]{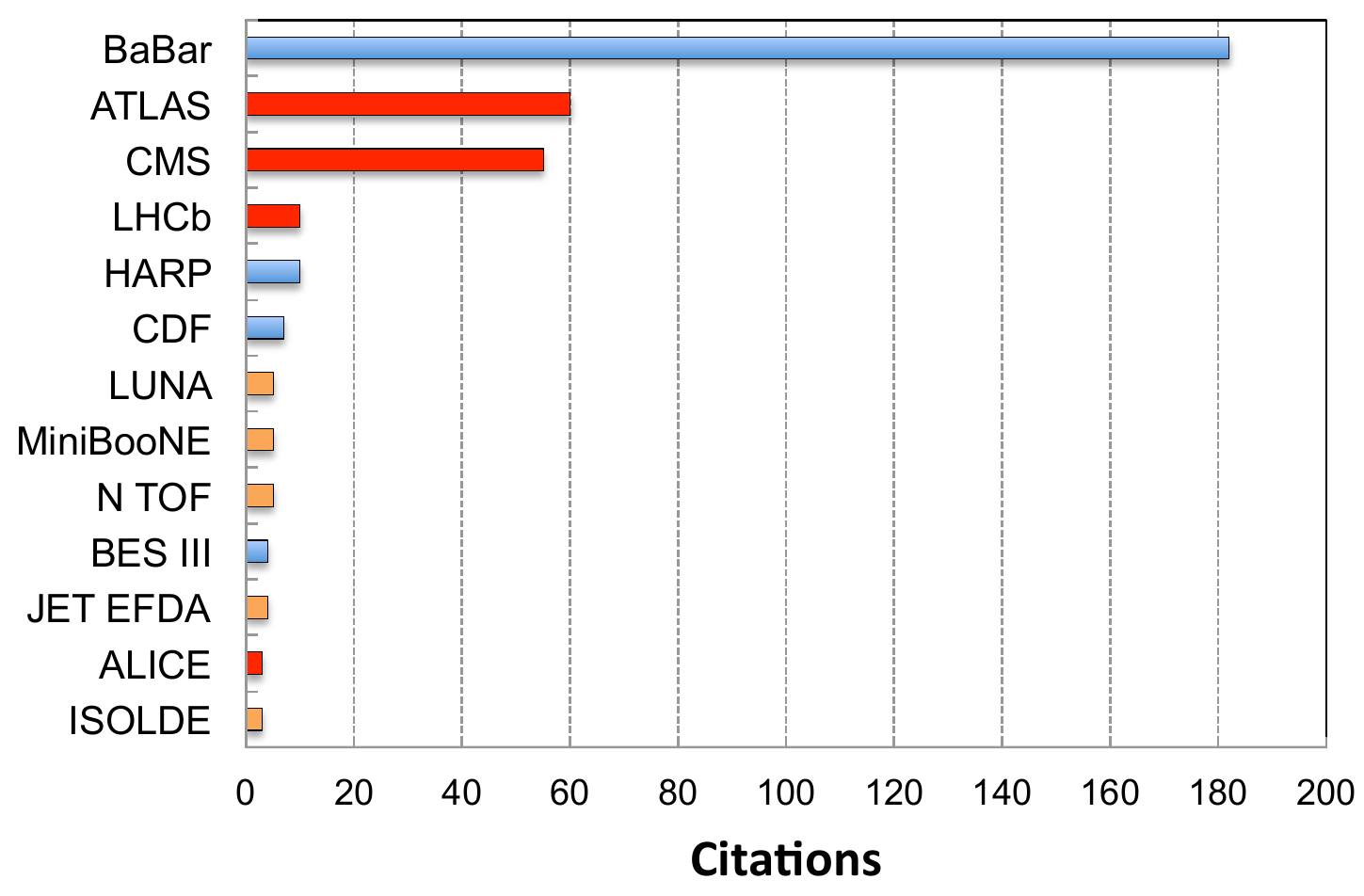}}
\caption{Experimental collaborations citing Geant4 reference \cite{g4nim}: LHC
experiments (red), other HEP experiments (blue), non-HEP experiments (yellow).
The collaboration listed in the histogram contribute approximately 16\% of the
total number of citations to \cite{g4nim}.}
\label{fig_g4cite_coll}
\end{figure}

One can observe in figure \ref{fig_g4cite_coll} that only a relatively small number
of citations to \cite{g4nim} are associated with LHC collaborations at this stage
of their life-cycle. 
It is worthwhile to note that only approximately 20\% of the citations to \cite{g4nim} 
listed in the Web of Science are formally associated with a collaboration (identified
as "Group Authors"); the vast majority of publications citing \cite{g4nim} appear as
the product of individual research groups, rather than of formal experimental 
organizations.
Figure \ref{fig_g4cite_coll} lists the collaborations that contribute the
largest number of citations; they correspond to approximately 75\% of the
citations associated with collaborations in the Web of Science.

ROOT is documented in two reference publications \cite{rootnim,rootcpc},
published in 1997 and 2009.
The earlier one is a contribution to a workshop proceedings, while the
later one is a regular journal publication.
These papers have collected respectively 540 and 27 citations (including
citations from conference proceedings indexed by the Web of Science).

\begin{figure}
\centerline{\includegraphics[angle=0,width=15cm]{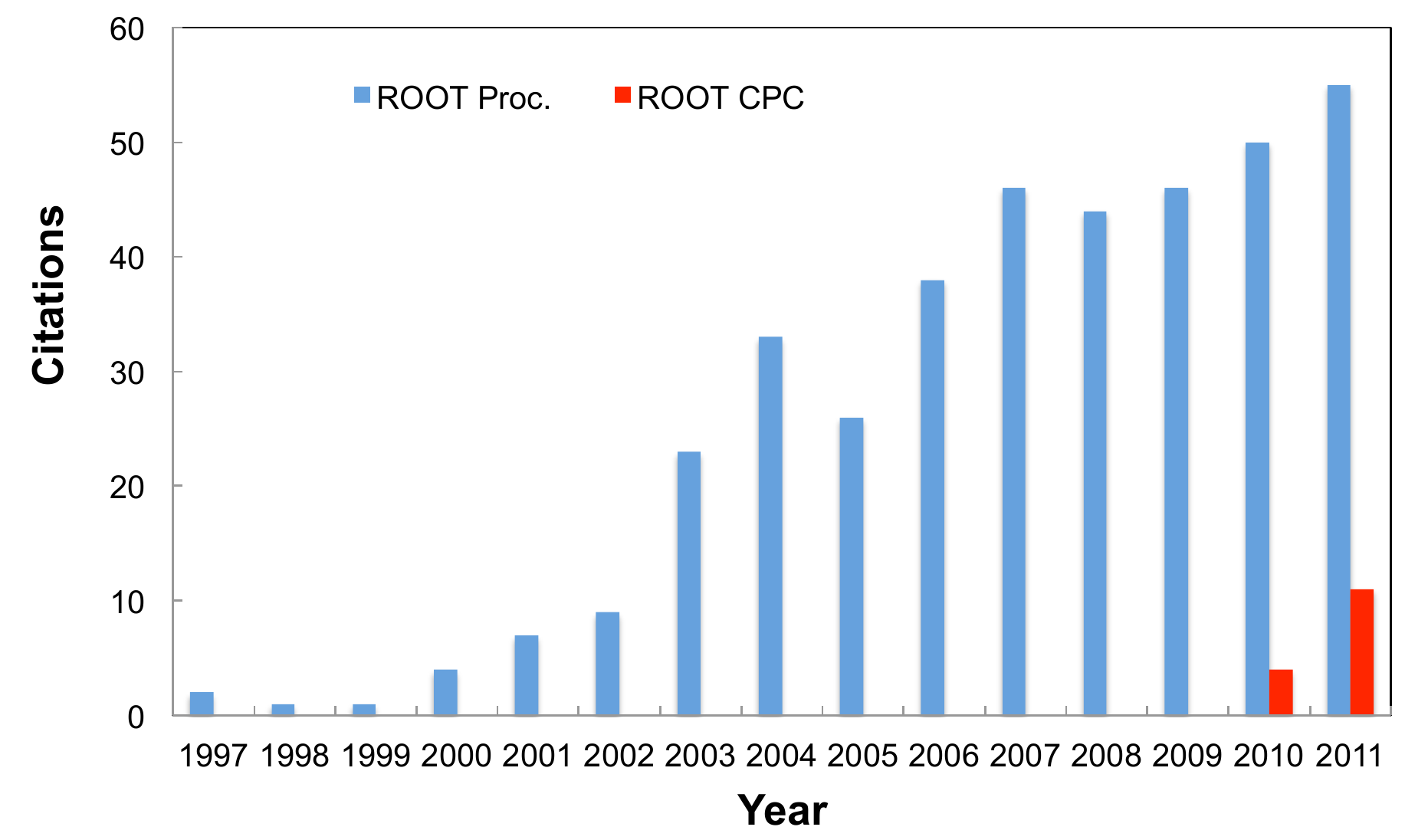}}
\caption{Number of citations collected by ROOT reference papers
\cite{rootnim,rootcpc} as a function of time.}
\label{fig_rootcite_years}
\end{figure}

The time distribution shown in figure \ref{fig_rootcite_years} shows a similar
pattern as in figure \ref{fig_g4cite_years}: the citation to the more recent
reference is omitted by most publications that cite the earlier one.

\begin{figure}
\centerline{\includegraphics[angle=0,width=16cm]{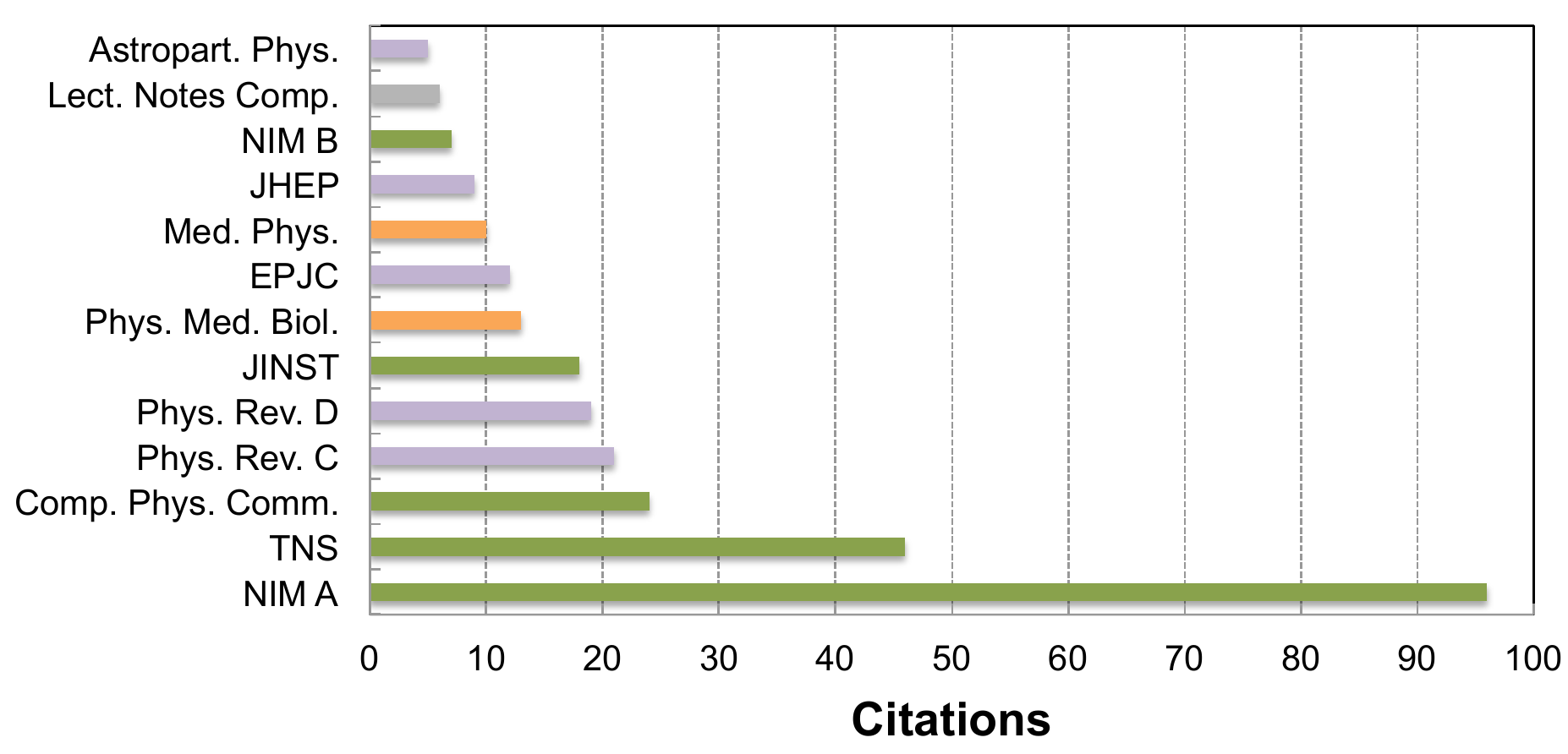}}
\caption{Journals citing ROOT reference \cite{rootnim}; the colour codes in the
plots are associated with the scope of the journals (physics: violet, nuclear
technology: green, medical physics: orange, radiation protection: light brown,
computing: grey. The journals listed in the histogram contribute approximately 75\% of the
total number of citations to \cite{rootnim}.}
\label{fig_rootcite_journals}
\end{figure}

The citations to ROOT earlier reference \cite{rootnim} have a multidisciplinary
character, as is visible in figure \ref{fig_rootcite_journals}, although the
relative contribution from various domains appears different for Geant4 and
ROOT.
The distribution of the domains of the citations listed in figures
\ref{fig_g4cite_journals} and \ref{fig_rootcite_journals} is summarized in table
\ref{tab_g4root}: citations to Geant4 appear equally distributed between physics
and nuclear technology journals, while nuclear technology journals are the most
relevant source of citations to ROOT;
also, the fraction of citations from medical physics and radiation protections
journals is significantly larger for Geant4 than for ROOT.
It is worhtwhile to remind the reader that table \ref{tab_g4root}, similarly to
figures \ref{fig_g4cite_journals} and \ref{fig_rootcite_journals}, reflects the
major sources of citations, amounting to approximately 75\% of total citations
collected by Geant4 and ROOT main references.

\begin{center}
\begin{table}[h]
\label{tab_g4root}
\caption{Source of citations to Geant4 and ROOT main reference papers.}
\centering
\begin{tabular}{l c c}
\hline
Journal categories & {\bf Geant4} 	& {\bf ROOT} \\
&  (\%) &  (\%) \\
\hline
Nuclear technology 	& 30 	& 50 \\
Physics	& 30	& 18 \\
Medical physics and radiation protection 	& 14		& 6 \\
\hline
\end{tabular}
\end{table}
\end{center}

\section{Publications by HEP experiments}
 
\begin{figure}
\centerline{\includegraphics[angle=0,width=10cm]{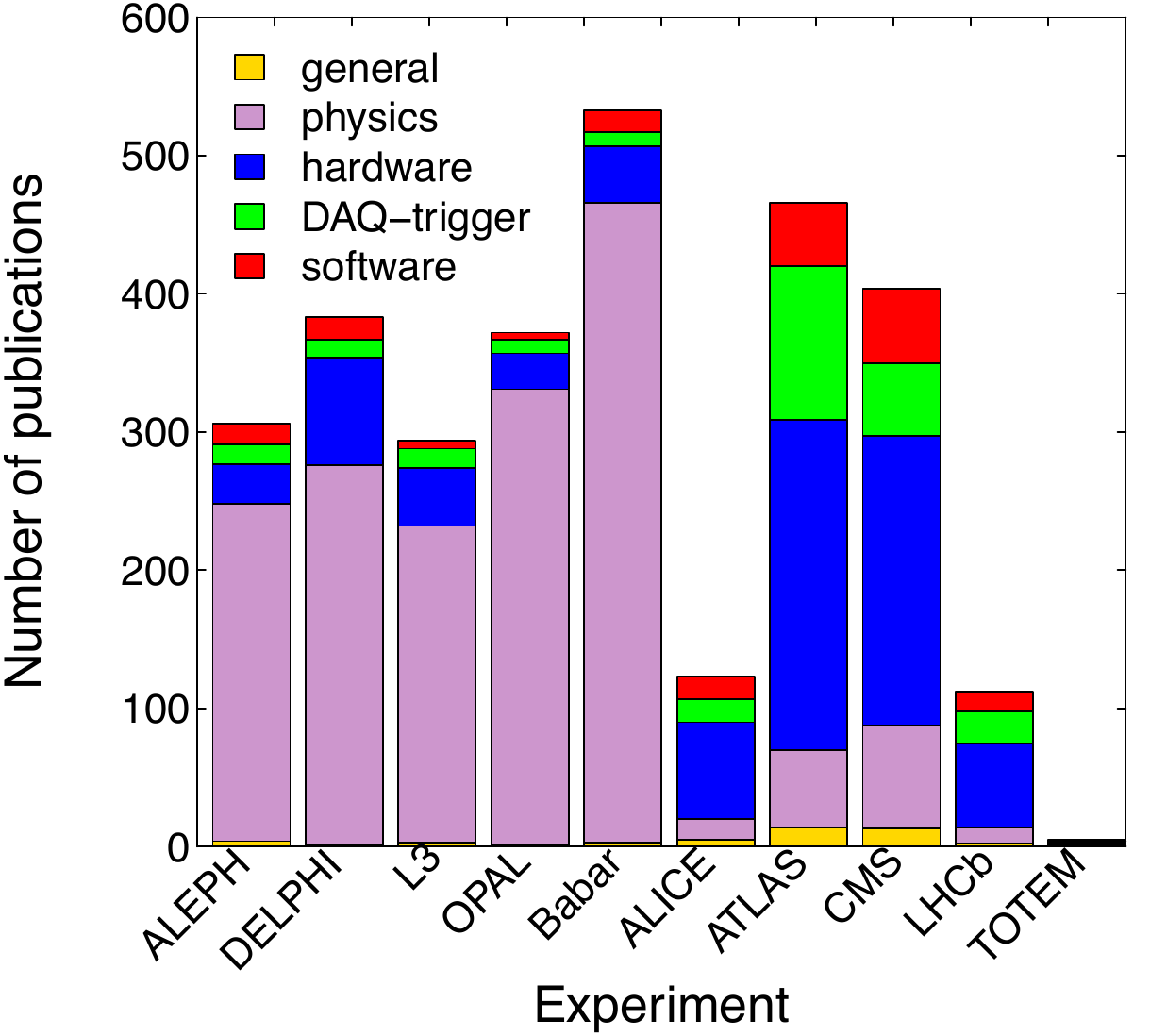}}
\caption{Papers published by the HEP experiments considered in this study,
distinguishing the contribution of various categories to the total count.}
\label{fig_exp_pub}
\end{figure}

The number of publications produced by the HEP experiments considered in this
study is plotted in figure \ref{fig_exp_pub}.
The plot distinguishes papers belonging to various categories: physics,
hardware, software, DAQ-trigger and general.
Physics papers are the dominant component for the experiment that terminated 
the data-taking phase and are close to the end of 
their lifecycle, while they represent a small fraction of the publications by 
LHC experiments, which are in the early stage of their run.
The last category includes papers describing the whole detector, or the
performance of some subsystems, which involve hardware and software aspects.

\begin{figure}
\centerline{\includegraphics[angle=0,width=8cm]{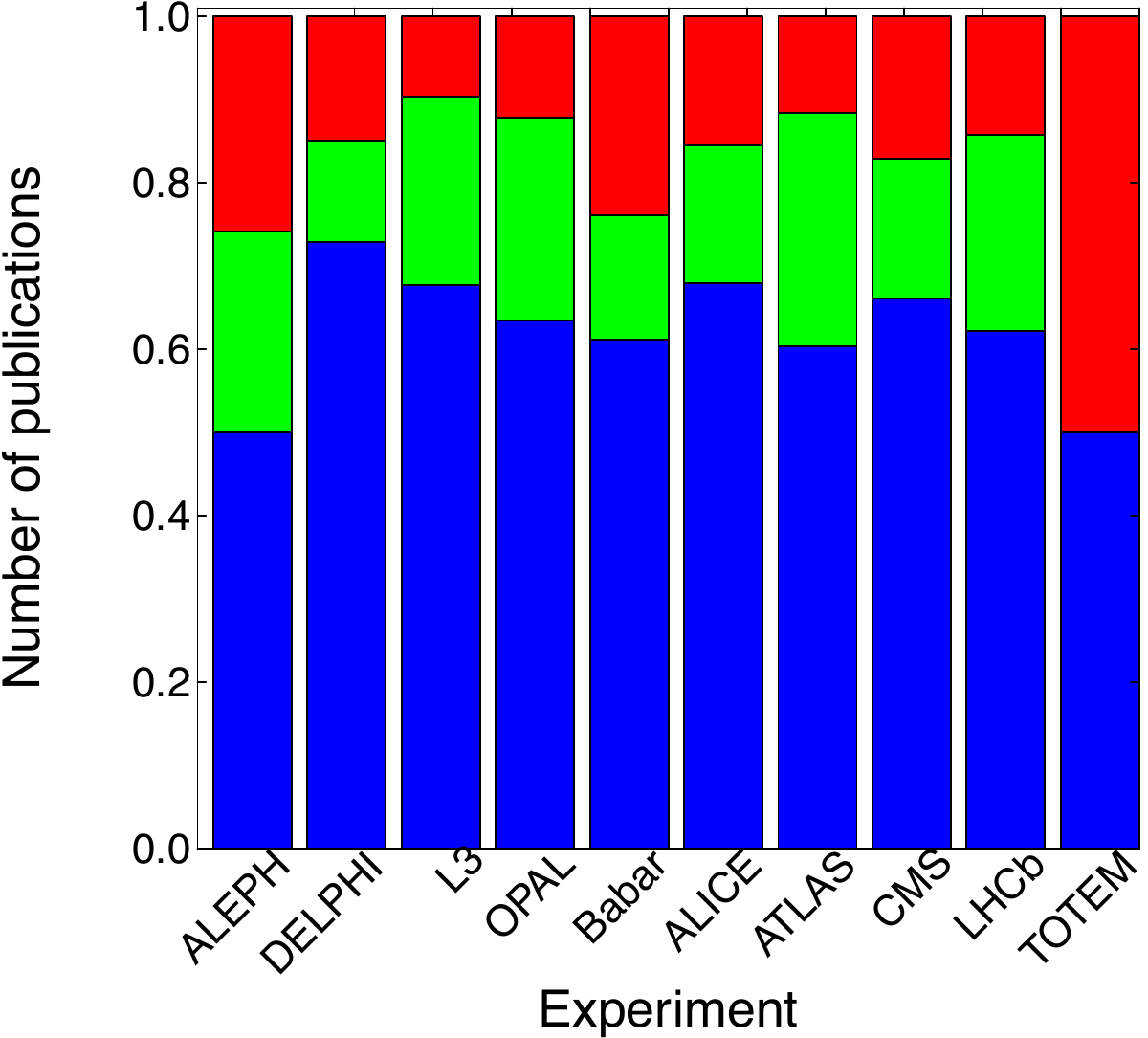}}
\caption{Technological papers published by the HEP experiments considered in this study:
fraction of hardware, software and DAQ-trigger publications.}
\label{fig_exp_pubtech}
\end{figure}

\begin{figure}
\centerline{\includegraphics[angle=0,width=16cm]{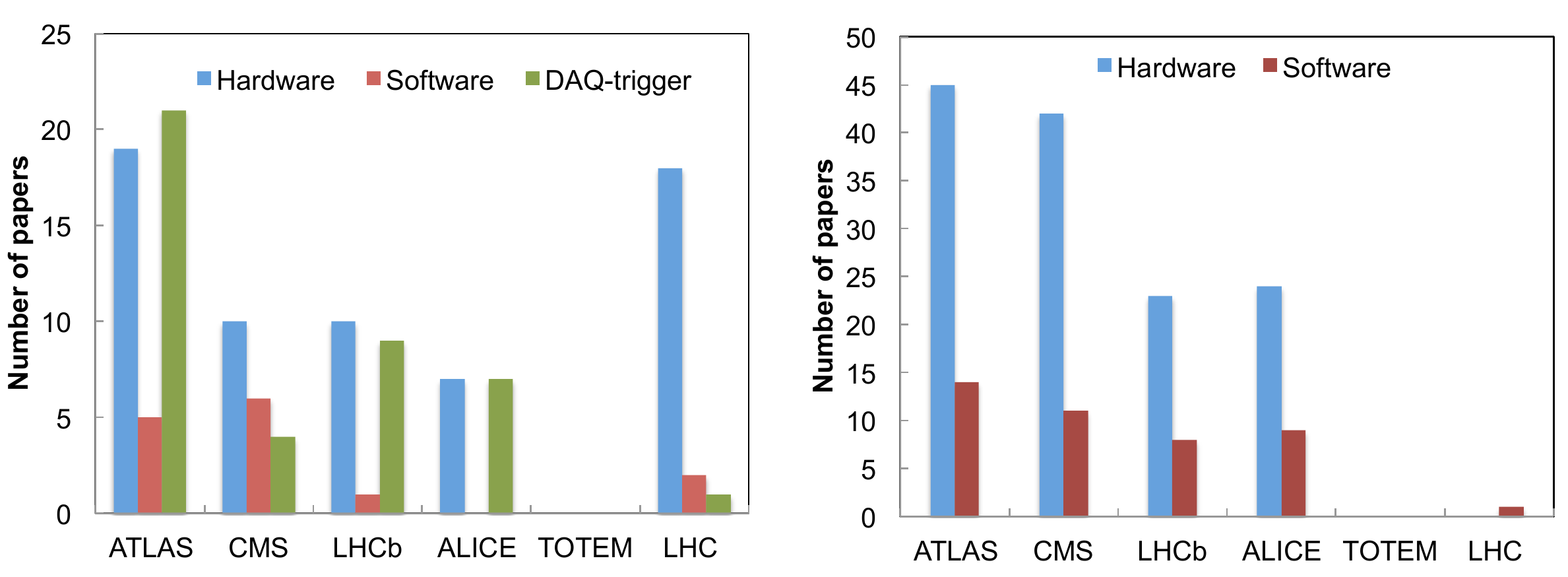}}
\caption{Technological papers published by the HEP experiments considered in this study
in TNS (left) and NIM A (right) since 2008:
hardware, software and DAQ-trigger publications. The bins identified 
as LHC correspond to papers related to LHC, but not specifically associated with 
any of the LHC experiments.}
\label{fig_exp_hwsw2008}
\end{figure}

Software related papers appear to be a small fraction of publications for all
the experiments: this trend is evident in figure \ref{fig_exp_pubtech}, which
shows the apportioning of technological papers across the three categories of
hardware, software and DAQ-trigger.
The relatively smaller presence of software publications in the production
of LHC experiments is confirmed in a more detailed analysis performed over
the papers published since the start of LHC operation in 2008 in two
representative nuclear technology journals, NIM A and TNS, shown in figure
\ref{fig_exp_hwsw2008}.

\begin{figure}
\centerline{\includegraphics[angle=0,width=8cm]{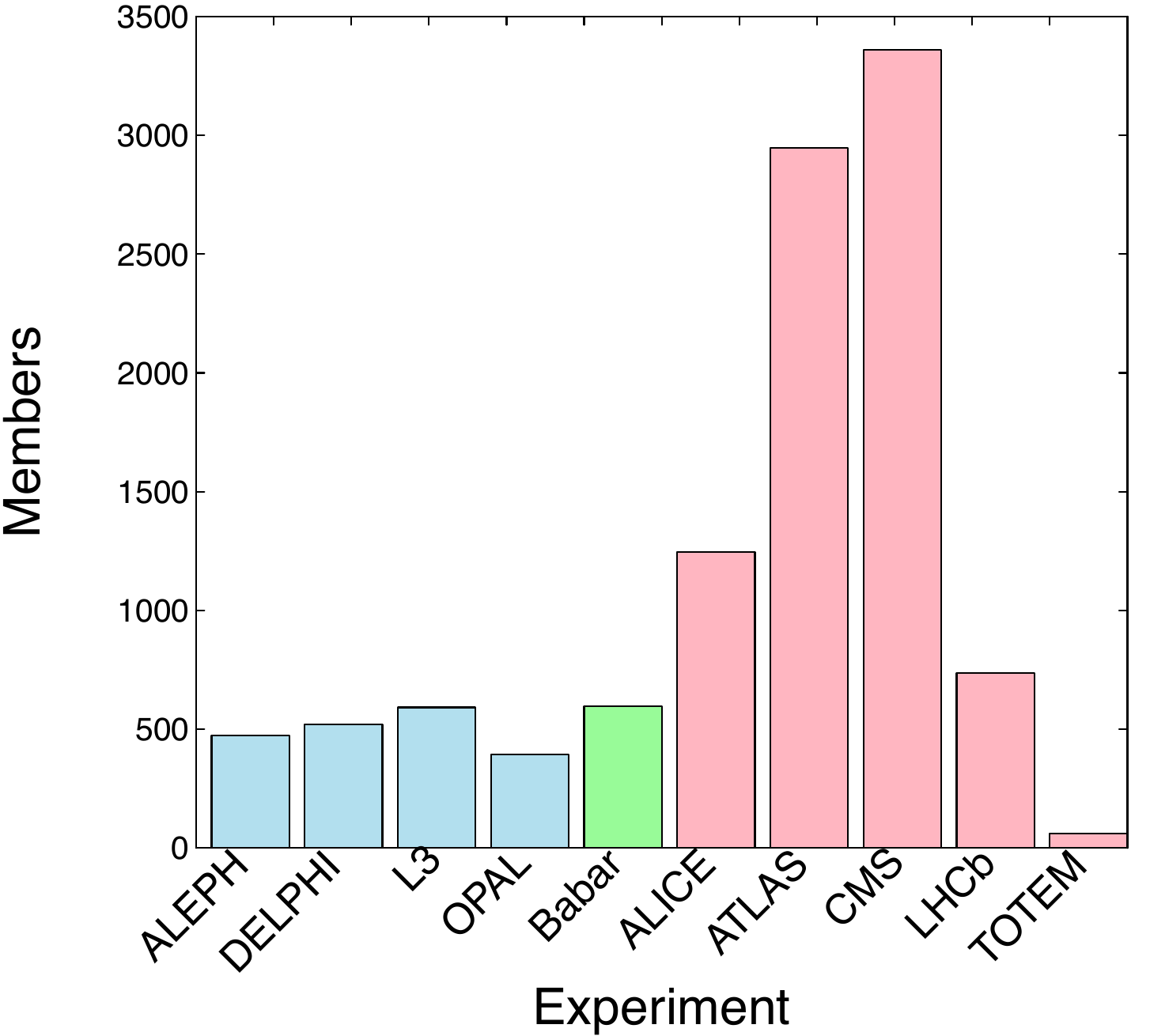}}
\caption{Size of the experimental collaborations considered in this study.}
\label{fig_members}
\end{figure}

The time distribution of the publications produced by the experiments considered
in this study is shown in figure \label{fig_exp_pubyear}.
The horizontal scale of the plots takes as a reference the year when LEP (1989),
BaBar (1998) and LHC (2008) started running.
Figure \label{fig_exp_pubyear} shows both the total count of papers produced per
year, and the number of published papers per collaboration member along the lifecycle of
the experiments.
The number of collaboration members is subject to variation over the lifetime of
an experiment, a constant number is assumed in this study, due to the difficulty
of ascertaining the number of collaboration members as a function of time for
all the experiments.
The size of the LHC collaborations is assumed to be the number of
members reported in CERN "Greybook" at the time of the CHEP conference;
for LEP experiments and BaBar the size of the collaboration was taken as
the number of authors of their most cited paper.
The number of collaboration members assumed in this study is shown in 
figure \ref{fig_members}.

\begin{figure}
\centerline{\includegraphics[angle=0,width=16cm]{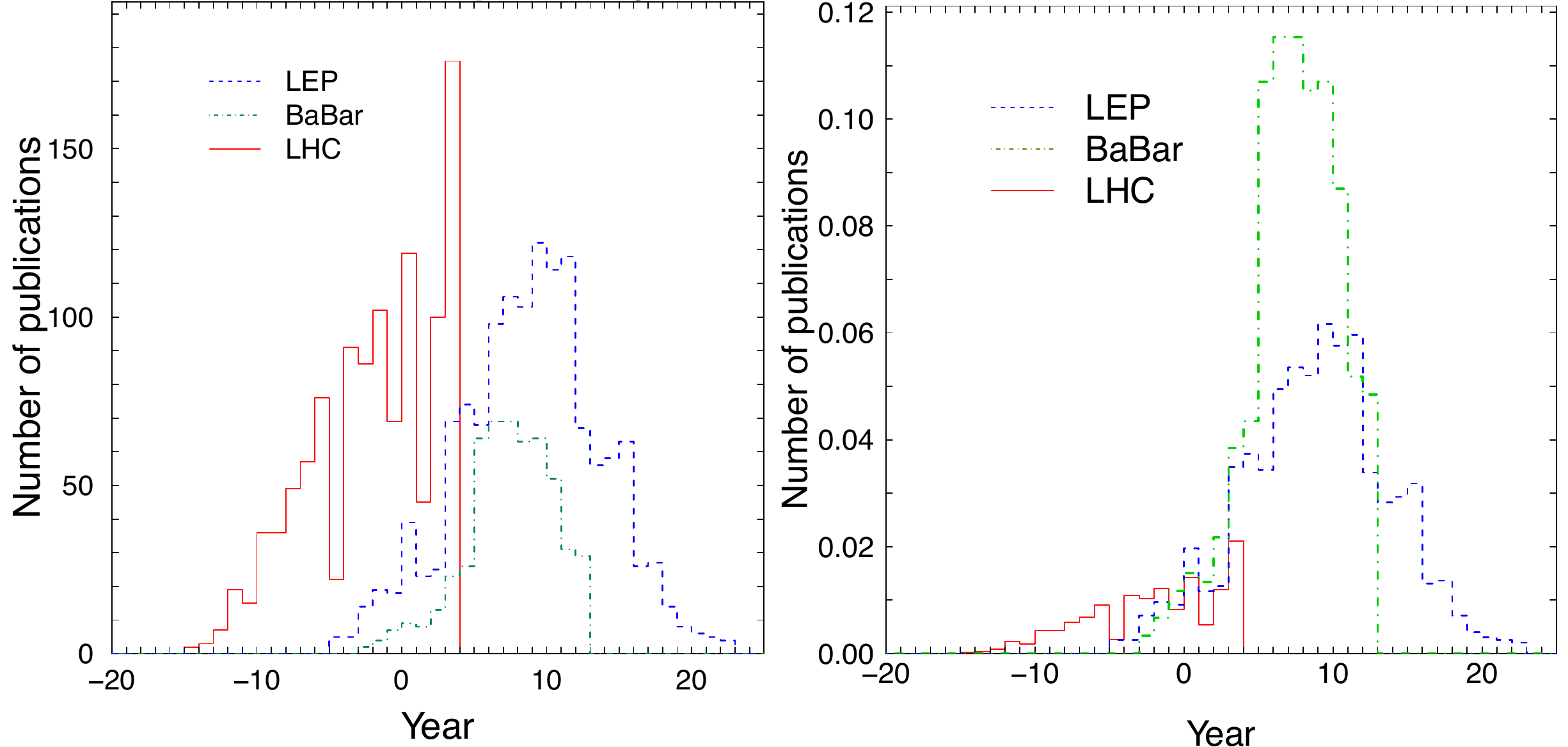}}
\caption{Papers published by the HEP experiments considered in this study 
as a function of operation year: total count (left) and number of published papers per
collaboration member (right). Years are counted with reference to the start of run
of LEP (1989), BaBar (1999) and LHC (2008).}
\label{fig_exp_pubyear}
\end{figure}

\begin{figure}
\centerline{\includegraphics[angle=0,width=8cm]{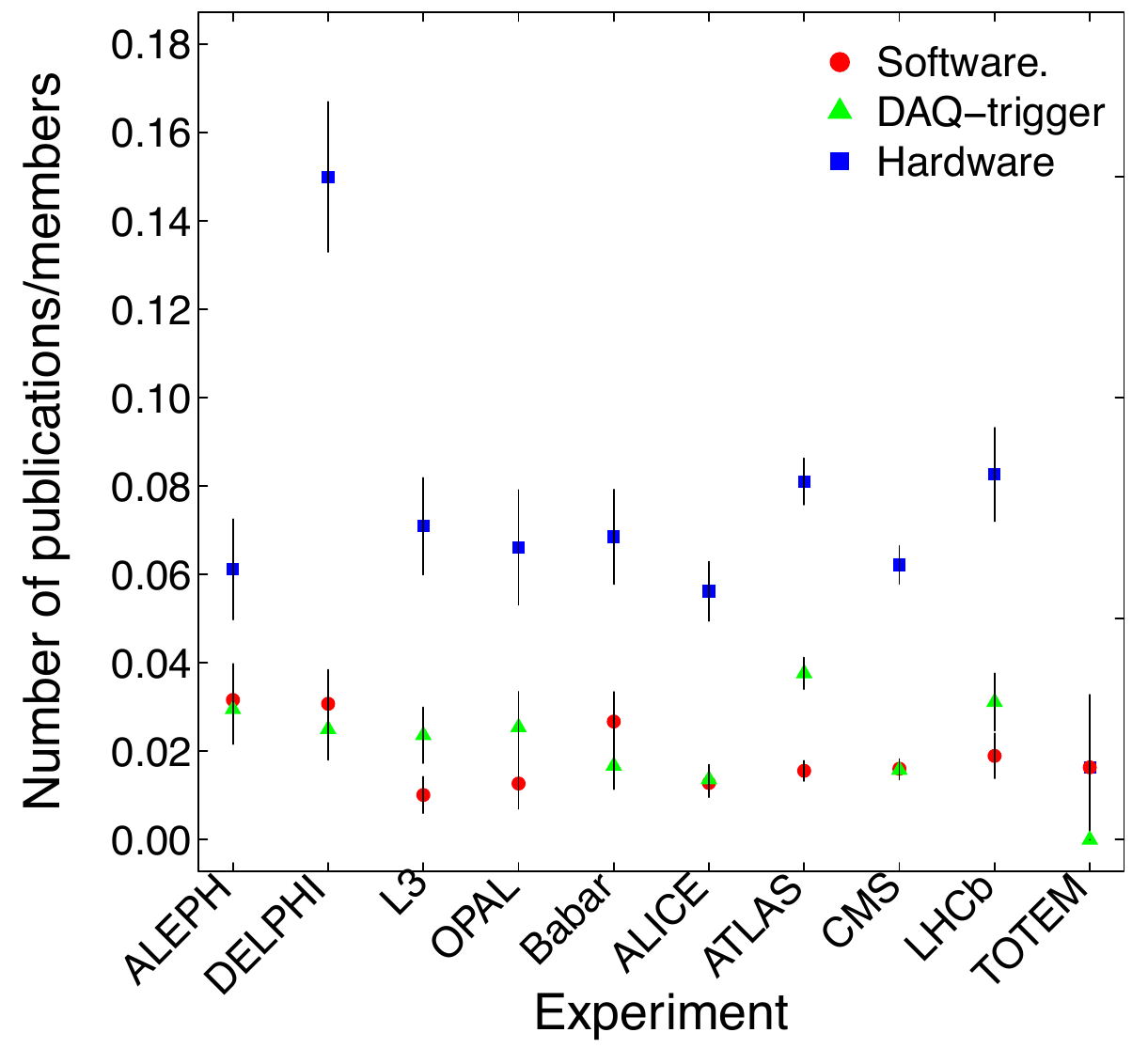}}
\caption{Number of hardware, software and DAQ-trigger papers published
by HEP experiments, scaled by the number of collaboration members.}
\label{fig_exp_techmem}
\end{figure}

\begin{figure}
\centerline{\includegraphics[angle=0,width=8cm]{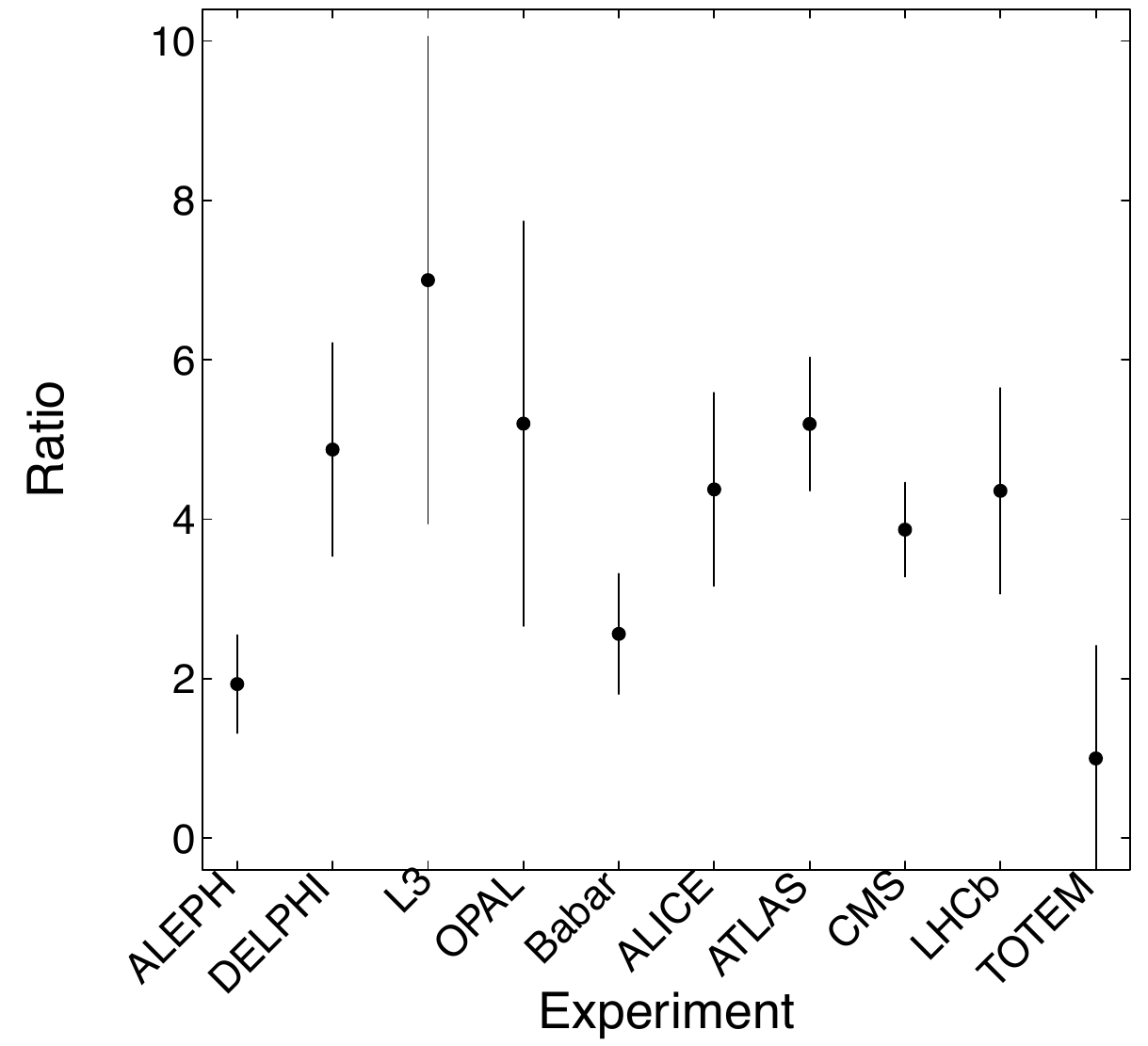}}
\caption{Ratio of hardware to software papers for a sample of HEP experiments.}
\label{fig_exp_hwsw}
\end{figure}

The number of hardware, software and DAQ-trigger publications appears
approximately constant of the three generations of HEP experiments considered in
this study, when it is scaled to the collaboration size, as shown in figure
\ref{fig_exp_techmem}.
The ratio of hardware to software publications, shown in figure
\ref{fig_exp_hwsw}, is also approximately constant across the experiments:
harware papers outnumber software ones by approximately a factor four.
This result differs from that reported in \cite{swpub}, which depicted an 
earlier stage of the lifecycle of LHC experiments, preceding the start of
LHC operation.
The difference could be also partly explained by evolutions in the Web of
Science since the publication of \cite{swpub}, namely the move of a large number
of conference papers to a dedicated database, which excludes them from the
analysis reported here.

\begin{figure}
\centerline{\includegraphics[angle=0,width=12cm]{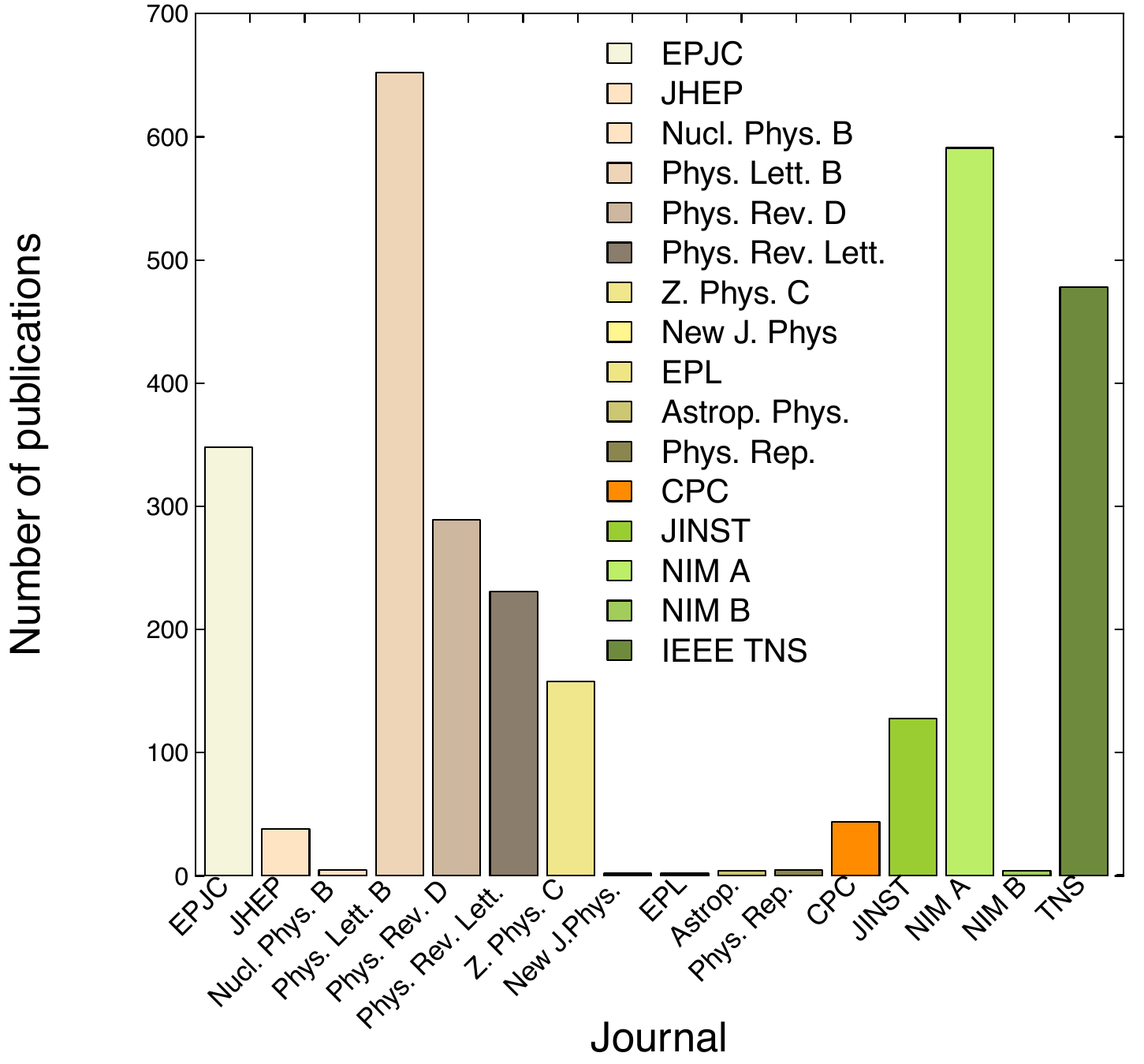}}
\caption{Journals where the HEP experiments considered in this study published their papers.}
\label{fig_journal_all}
\end{figure}

\begin{figure}
\begin{center}
\mbox
{
\subfigure[1982-1999]
{
\includegraphics[width=.5\textwidth]{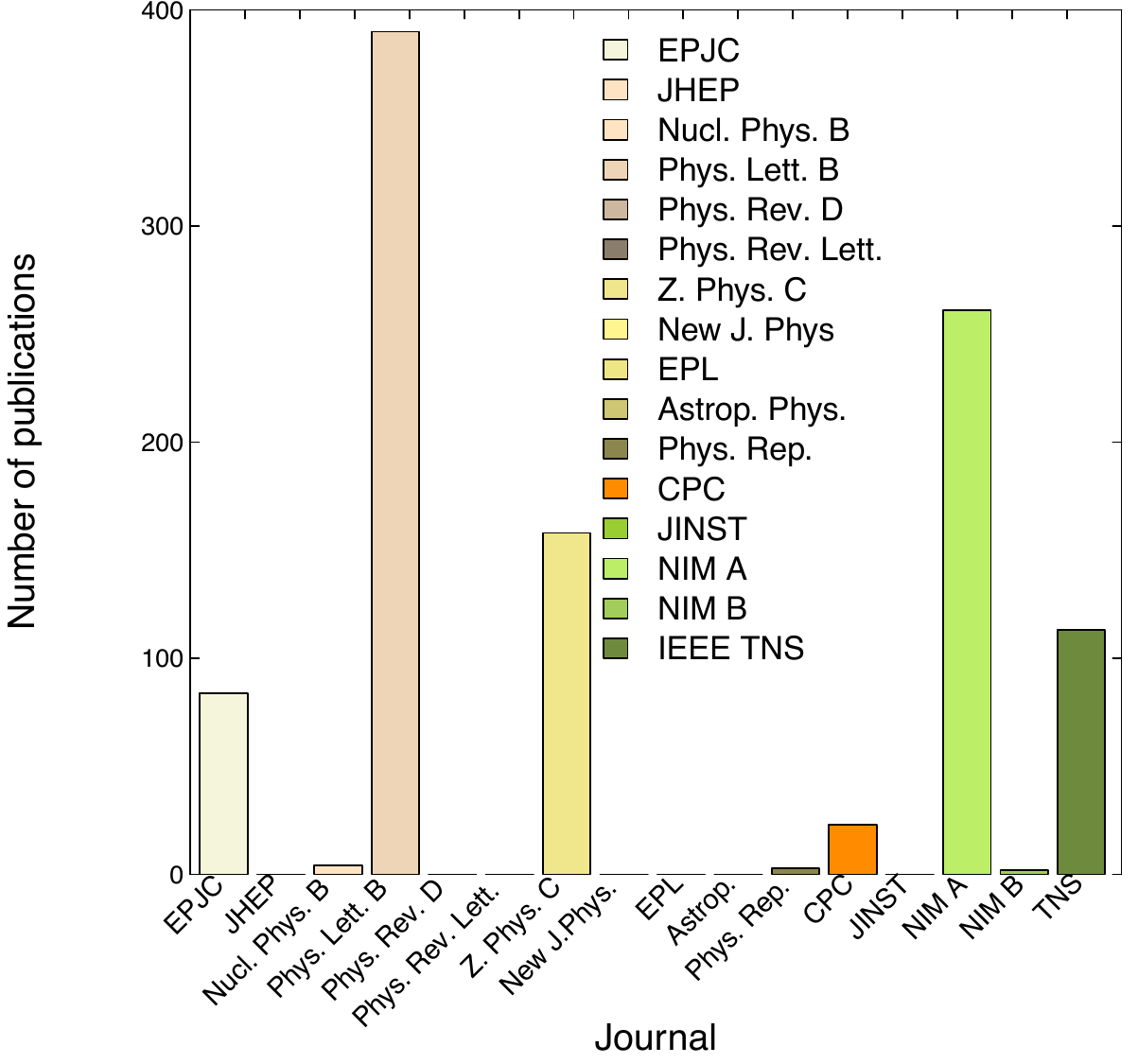}
\label{fig_ja}
}
\quad
\subfigure[2000-2011]
{
\includegraphics[width=.5\textwidth]{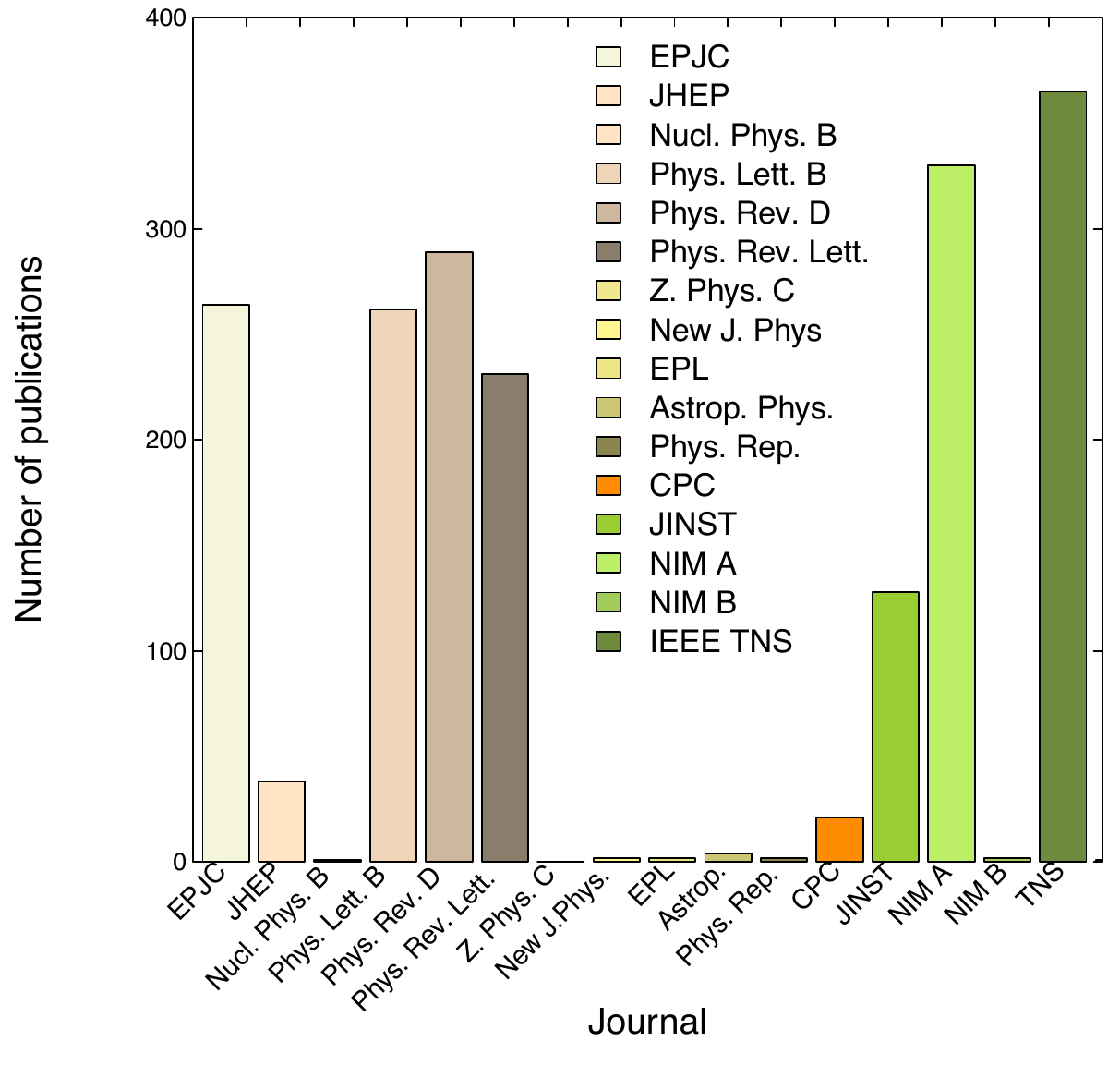}
\label{fig_jb}
}
}
\end{center}
\caption{Journals where HEP experiments published their papers: in years
1982-1999 (left) and 2000-2011 (right)}
 \label{fig_journal_years}
\end{figure}

Figure \ref{fig_journal_all} illustrates the distribution of papers published by HEP experiments
in physics and technological journals. 
The histogram involves the journals collecting the largest number of publications by
the experiments considered in this study.
One can observe in figure \ref{fig_journal_years} that the relative importance
of some journals has evolved over the years in the field: among technological
journals, TNS has increased its popularity in the HEP domain in the last decade,
while JINST (Journal of Instrumentation) is growing rapidly.

\begin{figure}
\centerline{\includegraphics[angle=0,width=14cm]{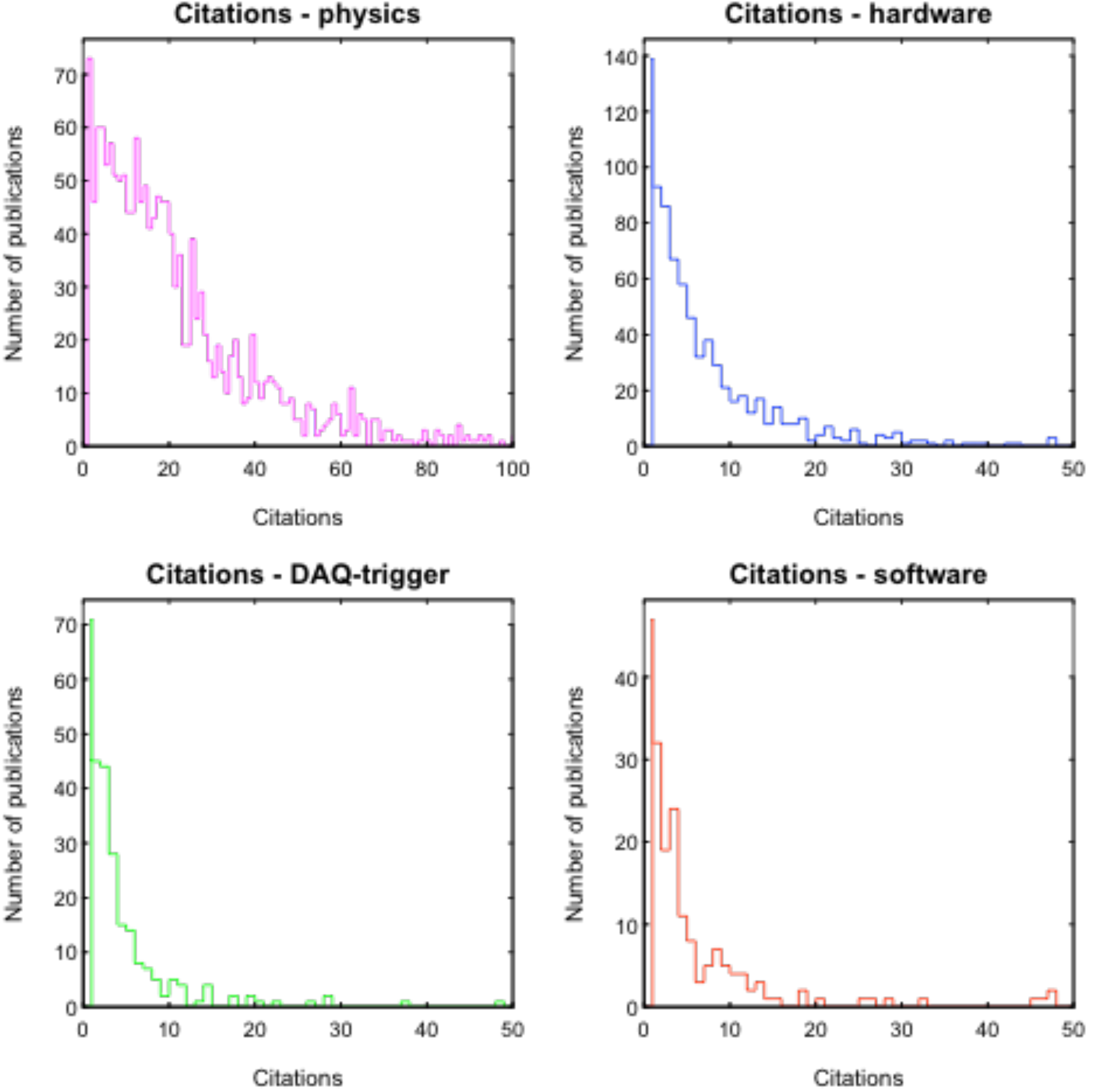}}
\caption{Number of citations to HEP experiments' publications: physics, hardware,
software and trigger/DAQ publications.} \label{fig_citations}
\end{figure}

\begin{figure}
\centerline{\includegraphics[angle=0,width=14cm]{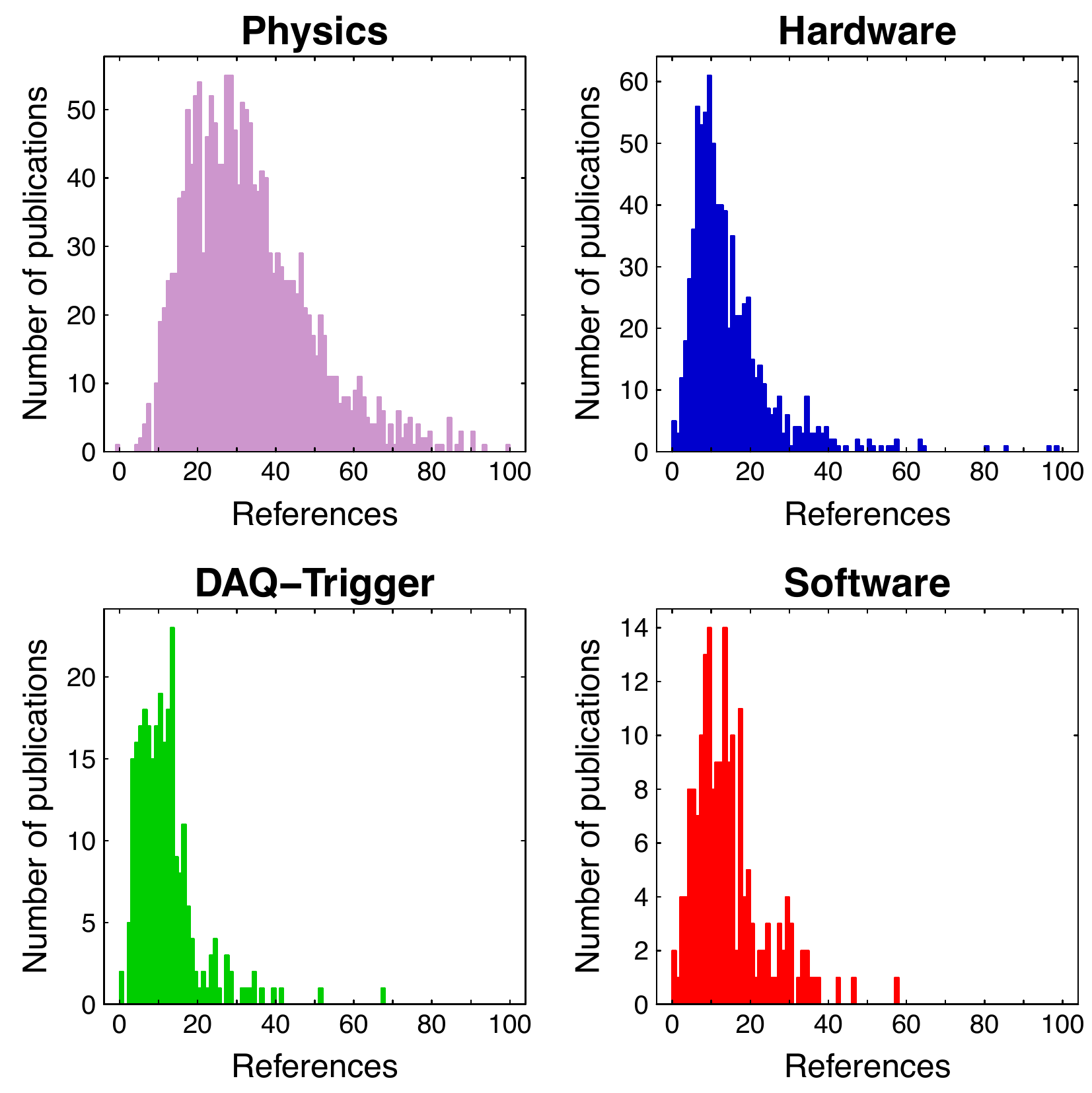}}
\caption{Number of references cited in HEP experiments' publications: physics,
hardware, software and trigger/DAQ publications.}
\label{fig_references}
\end{figure}

The distribution of the number of citations collected by various categories of
HEP experimental papers is shown in figure \ref{fig_citations}: physics papers
receive a larger number of citations than technological papers.
The fraction of physics papers that are not cited amounts to 4\%, while it is
much larger for technological papers: 17\% for hardware, 25\% for software and
27\% for DAQ-trigger publications within the data sample examined in this study.
Physics papers include a larger number of references than technological papers, as it
appears in figure \ref{fig_references}: the different citation habits in these two
domains are prone to affect the citation patterns shown in \ref{fig_citations}.

\begin{figure}
\centerline{\includegraphics[angle=0,width=16cm]{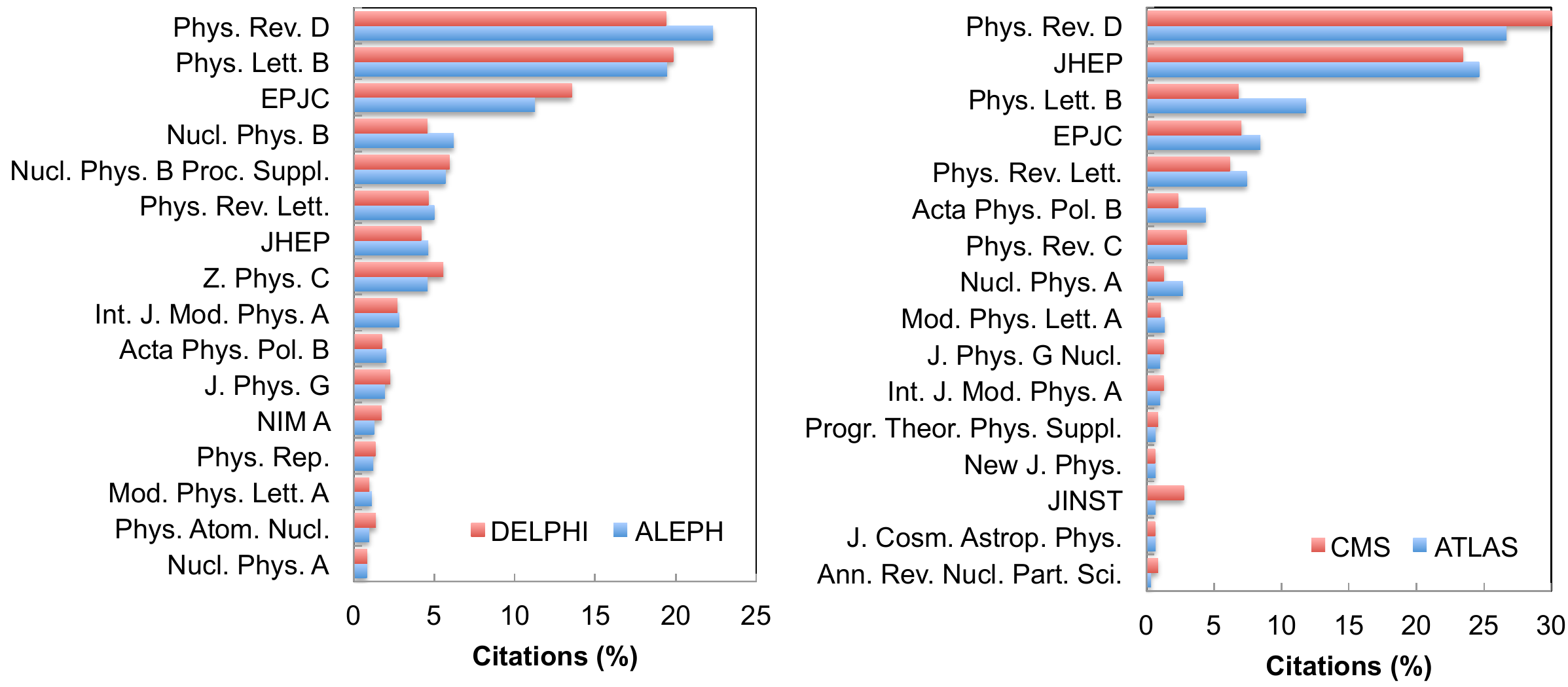}}
\caption{Sources of citations to physics papers published by representative LEP
experiments (left) and LHC experiments (right); the histograms include more than
90\% of the citations received by the physics papers of the selected
experiments.}
\label{fig_exp_physcite}
\end{figure}

\begin{figure}
\centerline{\includegraphics[angle=0,width=16cm]{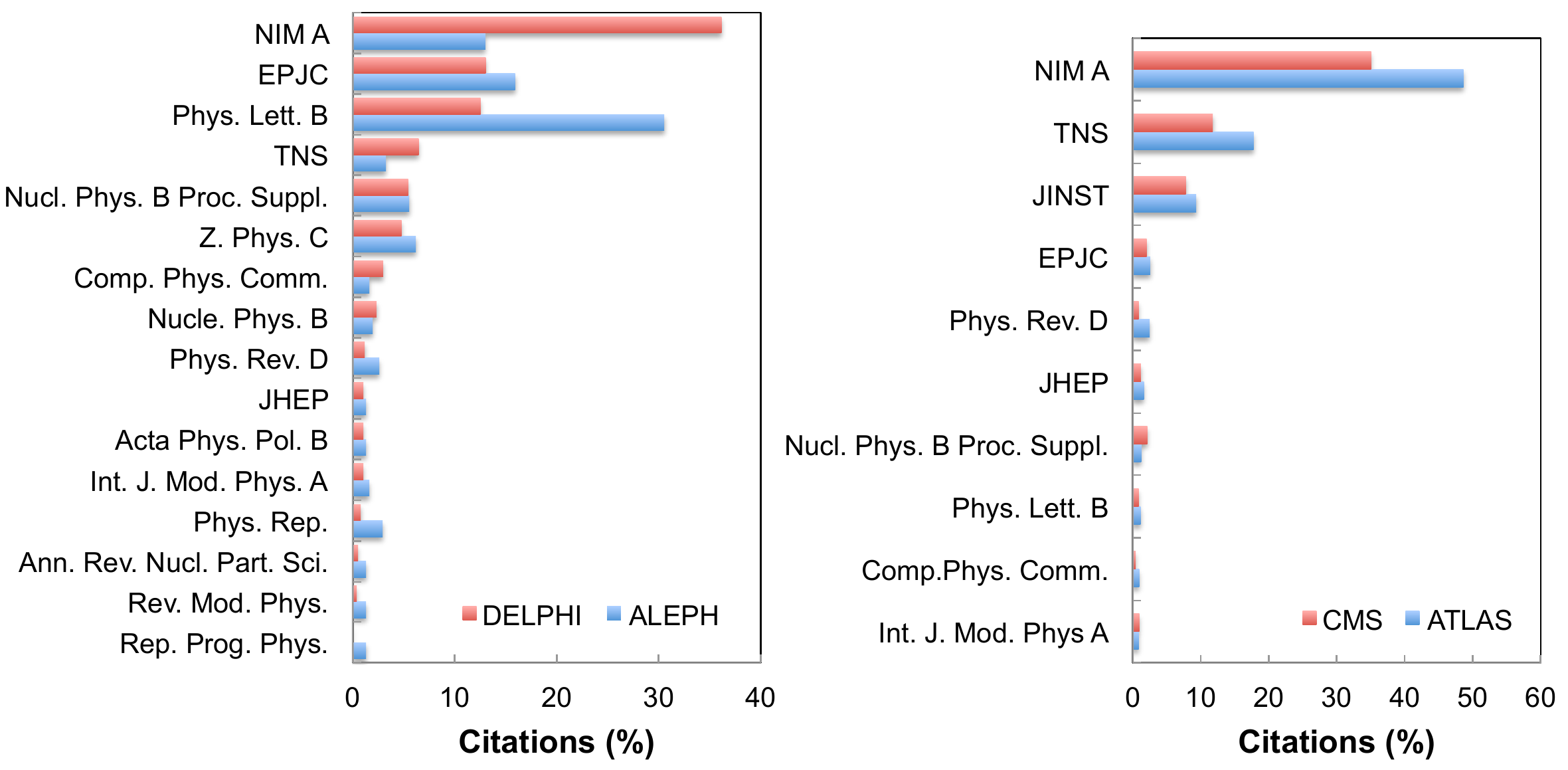}}
\caption{Sources of citations to technological papers published by
representative LEP experiments (left) and LHC experiments (right): the
histograms include more than 90\% of the citations received by the technological
papers of the selected experiments.}
\label{fig_exp_techcite}
\end{figure}

The citations to the physics papers of the HEP experiments considered in this
study come almost entirely from journals specialized in high energy physics or
closely related fields, such as nuclear physics and astroparticle physics:
the journals contributing more than 90\% of the citations to physics papers
published by representative LEP experiments (ALEPH and DELPHI) and
LHC experiments (ATLAS and CMS) are listed in figure \ref{fig_exp_physcite}.
Technological papers published by HEP experiments are cited by high energy and
nuclear physics journals, by nuclear technology journals and by review journals,
as is illustrated in \ref{fig_exp_techcite}.
Differently from what observed for the general software tools examined in section 
\ref{sec_tools}, the papers published by HEP experiments do not appear to
collect a significant number of citations from other disciplines, such as medical 
physics and radiation protection.

\begin{figure}
\centerline{\includegraphics[angle=0,width=16cm]{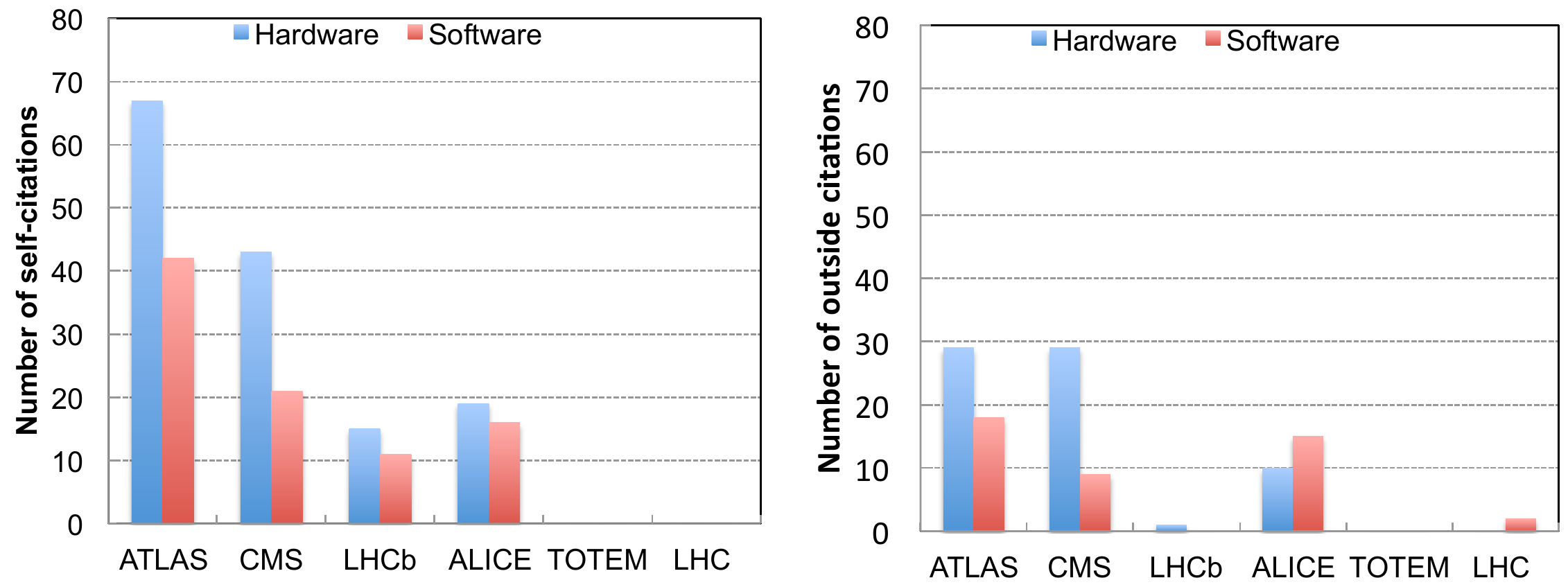}}
\caption{Number of self-citations (left) and outside citations (right) to papers
published by LHC experiments in NIM A since 2008: a citation is considered a
self-citation when the citing paper includes at least one of the authors of the
cited work, otherwise it is considered an outside citation. The bins identified 
as LHC correspond to papers related to LHC, but not specifically associated with 
any of the LHC experiments.}
\label{fig_exp_nimcite}
\end{figure}

\begin{figure}
\centerline{\includegraphics[angle=0,width=16cm]{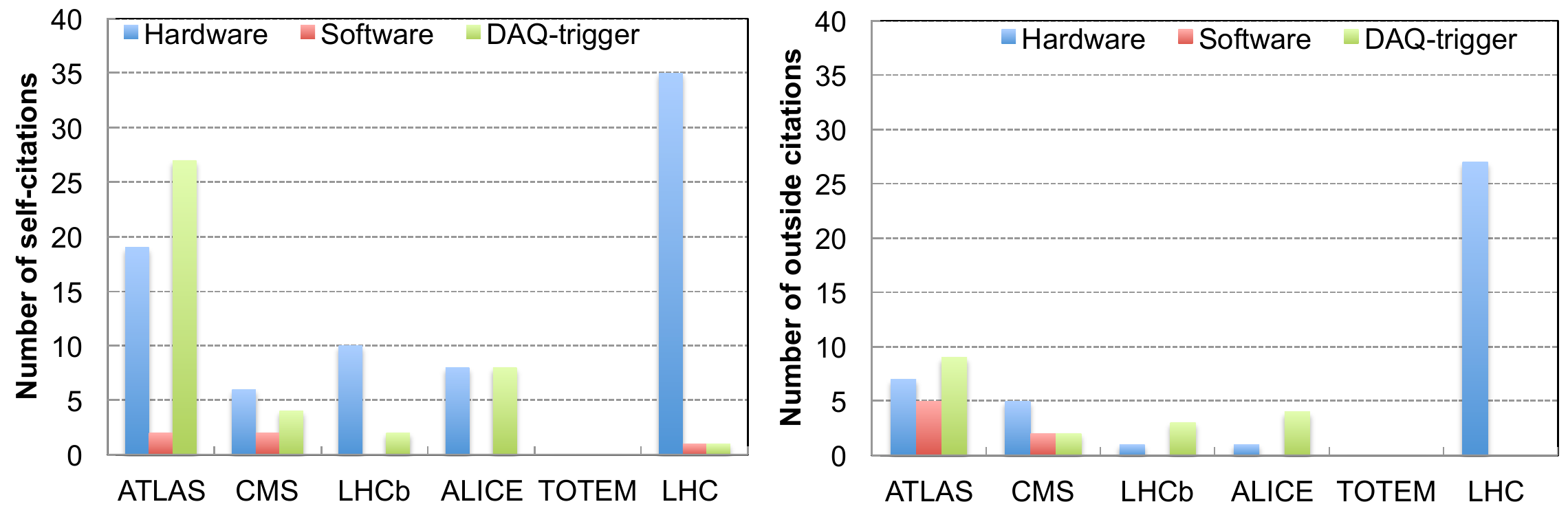}}
\caption{Number of self-citations (left) and outside citations (right) to papers
published by LHC experiments in TNS since 2008: a citation is considered a
self-citation when the citing paper includes at least one of the authors of the
cited work, otherwise it is considered an outside citation. The bins identified 
as LHC correspond to papers related to LHC, but not specifically associated with 
any of the LHC experiments.}
\label{fig_exp_tnscite}
\end{figure}

A large fraction of the citations collected by LHC technological publications 
consists of self-citations (i.e. the citing papers include at least one of the authors
of the cited work): this pattern is illustrated in figures \ref{fig_exp_nimcite} and
\ref{fig_exp_tnscite} for the two journals collecting the largest number of
technological publications by LHC experiments, NIM A and TNS.

The most cited publications produced by HEP experiments are in most cases the
respective reference papers describing the whole detector: these papers are
usually cited by the papers reporting the physics results of the experiment.
In the data sample examined in this study, excluding LHC experiments,  
the number of citations collected by the most cited paper varies from 309 for
the DELPHI experiment to 859 for the BaBar experiment; the citation statistics is
not yet meaningful for LHC experiments, that are at an early stage of their 
physics production.

\section{Publications related to the LHC Computing Grid}

Grid computing is an essential component of the operation of LHC experiments: a
large effort has been invested in the past decade to develop the grid computing
infrastructure and several application tools used by LHC experiments within a
project known as ``LHC Computing Grid'' (LCG).

A search for publications associated with LCG in the Web of Science results in a
small sample, consisting of less than 20 papers.
Grid computing has represented a large fraction of the scientific program of the
CHEP conference for the past decade, in addition to dedicated conferences. 
The small sample of journal publications related to LCG retrieved in the Web of
Science suggests that only a limited fraction of conference presentations in
this field evolves into regular publications in scholarly journals.

Due to the small sample size, a statistical analysis of LCG publications does 
not appear meaningful.

\section*{Conclusions}

The scientometric analysis reported in this paper provides a quantitative
overview of publication patterns in HEP experiments, covering the last three
decades.

The analysis has confirmed the general trend observed in a previous study:
software related papers are largely underrepresented with respect to 
hardware papers in the high energy physics experimental environment.
The ratio of hardware to software papers is approximately constant over
the experiments of the LEP and LHC generations.

Software papers collect in average fewer citations than hardware papers (and
physics papers); they also cite fewer references in their bibliography.

The analysis of citations to papers published by HEP experiments shows 
that both physics and technological papers collect the largest number of citations within
the HEP environment; a small fraction of citations comes from closely related
fields, such as nuclear and astroparticle physics.

General software tools motivated by the requirements of HEP experiments, such as
Geant4 and ROOT, exhibit different patterns.
The earlier Geant4 reference \cite{g4nim} has received more than 3000 citations
at the time of writing this paper: it is a landmark paper in Thomson-Reuters
Nuclear Science and Technology category, and the most cited publication for
major institutions such as CERN and INFN.
The analysis of the citations collected by these software tools shows 
the multidisciplinary character of these tools, which appear to be used in 
a variety of experimental fields not limited to HEP.
Geant4 is cited by a large number of physics papers, which confirm its significant 
role in the production of physics results by HEP experiments in the LHC era.

\section*{Acknowledgments}
The authors are grateful to the CERN Library for the support provided to this study.

\end{document}